\documentclass[aps,showpacs,twocolumn]{revtex4-1}

\usepackage{graphicx}
\usepackage{dcolumn}
\usepackage{bm}
\usepackage{float}
\usepackage{amssymb}

\usepackage{amsmath,amsthm}
\usepackage{xspace}
\usepackage{bm}

\begin{document}
\newcommand{\gSpinMix}{\ensuremath{g_{\uparrow\!\downarrow}}\xspace}
\newcommand{\VISHE}{\ensuremath{V_\mathrm{ISHE}}\xspace}
\newcommand{\Js}{\ensuremath{\bm{J}_\mathrm{s}}\xspace}
\newcommand{\sigmaN}{\ensuremath{\sigma_{\mathrm{N}}}\xspace}
\newcommand{\alphaSH}{\ensuremath{\alpha_{\mathrm{SH}}}\xspace}
\newcommand{\lambdaSD}{\ensuremath{\lambda_{\mathrm{SD}}}\xspace}

\title{Experimental observation of a large ac-spin Hall effect}

\author{Dahai Wei}

\author{Martin Obstbaum}
\author{Christian Back}
\author{Georg Woltersdorf}
\affiliation{Institut f{\"{u}}r Experimentelle und Angewandte Physik, Universit{\"{a}}t Regensburg, Universit{\"{a}}tsstra{\ss}e 31, 93053  Regensburg, Germany.}

\date{\today}

\begin{abstract} \bf
In spinelectronics the spin degree of freedom is used to transmit and store information. Ideally this occurs without net charge currents in order to avoid energy dissipation due to Joule heating. To this end the ability to create pure spin currents i.e.~without net charge transfer is essential. Spin pumping is the most popular approach to generate pure spin currents in metals \cite{tserkovnyak2002enhanced,urban2001gilbert,mizukami2001study,mosendz2010quantifying,azevedo2011spin}, semiconductors \cite{Ando2011NatMa,chen2013direct}, graphene \cite{PhysRevB.87.140401}, and even organic materials \cite{Ando2013NatMa}. When the magnetization vector in a ferromagnet (FM) - normal metal (NM) junction is excited the spin pumping effect leads to the injection of pure spin currents in the normal metal. The polarization of this spin current is time dependent \cite{Tserkovnyak-PRL02} and contains a very small dc component \cite{Brataas-spinbat03}.  The dc-component of the injected spin current has been intensely studied in recent years and has given rise to controversial discussions concerning the magnitude the spin Hall angle which is a material dependent measure of the efficiency of spin to charge conversion \cite{PhysRevB.83.174405,liu2012spin}. However in contrast to the rather well understood dc component \cite{mosendz2010quantifying,azevedo2011spin,saitoh2006conversion} the two orders of magnitude larger ac component has escaped experimental detection so far \cite{Jiao-PRL13}.  Here we show that the large ac component of the spin currents  can be detected very efficiently using the inverse spin Hall effect (ISHE).  The observed ac-ISHE voltages are one order of magnitude larger than the conventional dc-ISHE measured on the same device. The spectral shape, angular dependence, power scaling behavior and absolute magnitude of the signals are in line with spin pumping and ISHE effects. Our results demonstrate that FM-NM junctions are very efficient sources of pure spin currents in the GHz frequency range and we believe that our result will stimulate the emerging field of ac spintronics \cite{Kochan-PRL11,Jiao-PRL13}.

\end{abstract}


\maketitle


\begin{figure*}[!ht] \center \includegraphics[width=0.85\textwidth]{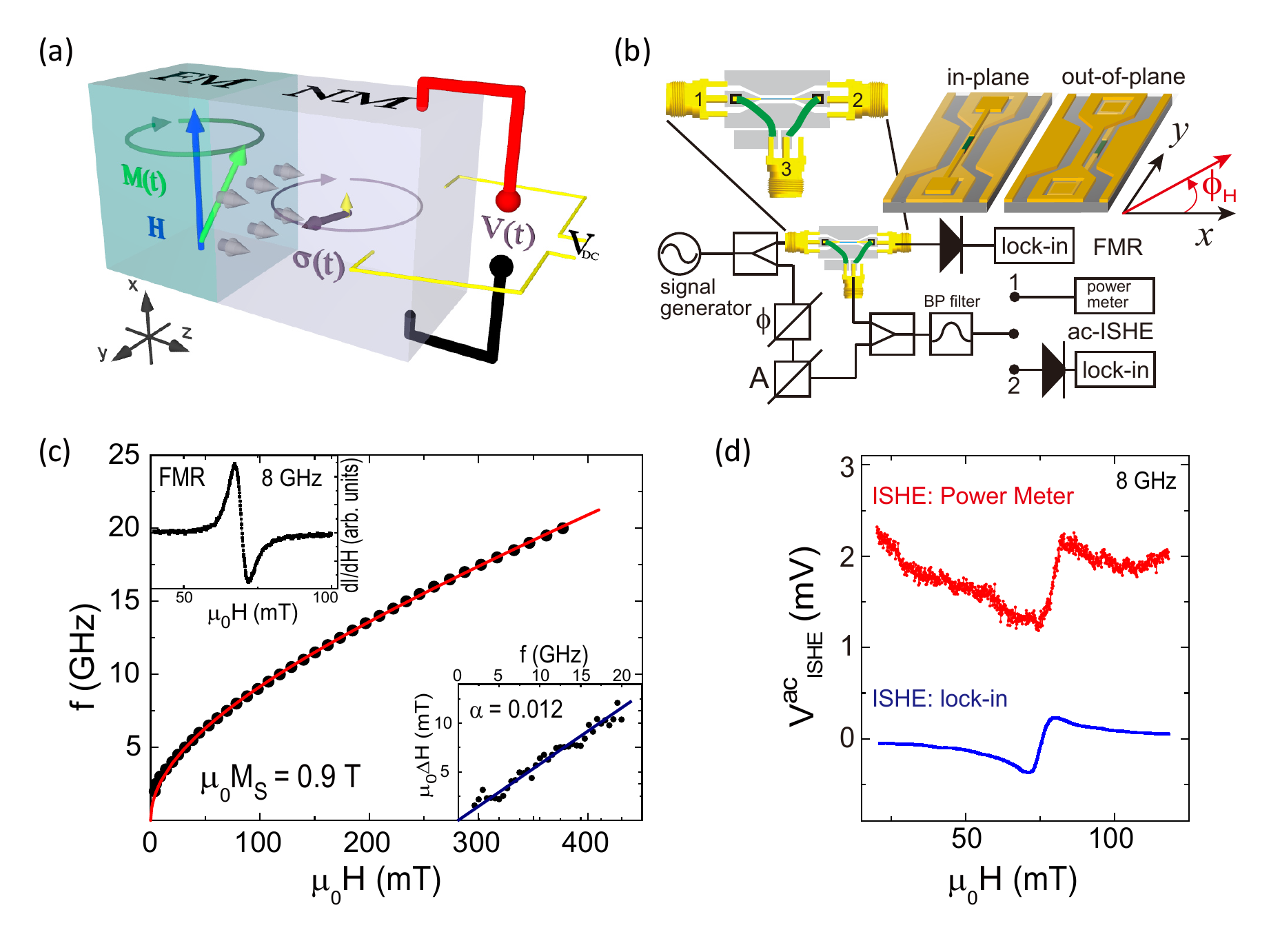}
 \caption
{\label{fig1} \textbf{Detection of ac-spin currents by ISHE.} (a) Illustration of spin currents due to ferromagnetic resonance based spin pumping and its detection using the inverse spin Hall effect. The time dependent spin polarization (indicated as purple arrow) rotates almost entirely in the $y-z$ plane. The small time averaged dc component (yellow arrow) appears along the $x$ axis. Both components can be converted into ac and dc voltages in the Pt layer along the $x$ and $y$ direction, respectively. (b) Layout of the measurement configuration. The rf signal is split into a part that excites the sample and a reference arm where amplitude and phase can be adjusted independently. The signal on terminal 2 is used for inductive FMR measurements, while the signal on terminal 3 originates from  ac-ISHE. This signal is either measured using a power meter or a lock-in amplifier. In-plane rf-excitation ($h_y$) is used when the bilayer stripe is placed on top of the signal line of the CPW, while placing the bilayer in the gap between signal line and ground planes leads to an out-of-plane excitation field ($h_z$). (c) FMR resonance field as a function of  microwave frequency. The upper left inset shows a typical FMR spectrum of the Pt/Ni$_{80}$Fe$_{20}$ bilayer measured at 8~GHz, the  bottom right inset is the frequency dependence of the resonance line width $\mu_{0}\Delta\rm{H}$. (d) Ac-ISHE spectra at 8~GHz measured using a power meter (red) and measured using field modulation and lock-in amplification (blue).
} \label{fig1}
\end{figure*}

The time dependence of the polarization of a spin current injected by spin pumping is given by $\bm{\sigma} \sim \bm{m}\times d\bm{m}/dt$ \cite{Tserkovnyak-PRL02} and is illustrated in Fig.~1(a).
The absorption of a spin current in a nonmagnetic metal with a finite spin Hall effect leads to an electric field $\bm{E}$ and therefore a voltage transverse to the spin current $\Js$ and spin polarization $\bm{\sigma}$ as expressed by the following equation:
\begin{eqnarray}
V_{\rm{ISHE}} \sim \bm{E}  \sim \Js \times \bm{\sigma} ,  \label{she}
\end{eqnarray}
where $\Js$ is the spin-current propagation direction, and $\bm{\sigma}$ denotes the spin polarization vector of the spin current. In this way dc and ac voltage signals may be measured as shown in  Fig.~1(a).
In the following we demonstrate experimentally the presence of a large ac component in the ISHE voltage signal in a FM/NM bilayer, where the ac spin current is generated by spin pumping at ferromagnetic resonance (FMR). The magnitude of the ac-ISHE signal is measured as a function of frequency, angle and power. In addition the dc and ac-ISHE signals are measured in the same device in order to compare their relative amplitudes.

A sketch of the experimental configuration is shown in Fig.~1(b), the FM/NM bilayer stripes are either integrated on top of the signal line or in the gap between the signal and ground lines of a coplanar waveguide (CPW). In these two configurations the magnetization in the FM is excited by  an in-plane and out-of-plane microwave magnetic field $\bm{h}_{\rm rf}$, respectively. In all experiments the stripes are $10~\mu$m wide, $400~\mu$m long, and the thicknesses of the Ni$_{80}$Fe$_{20}$ and NM layers are 10~nm.

The difficulty to detect the ac-ISHE signal lies in the ability to measure sub-mV GHz signals and isolate them from a large background signal caused by the excitation of FMR which has the same frequency.
The experimental setup for the measurement of ferromagnetic resonance (FMR) and ac-ISHE is sketched in Fig.~1(b).  The microwave signal transmitted from terminal 1 to terminal~2 is used to measure FMR inductively. In order to measure ac-ISHE signals the NM/FM stripe is connected to a 50~$\Omega$ waveguide (terminal~3). In addition the sample structure was designed as waveguide (strip line) such that the ac-ISHE voltage signal can propagate along the NM/FM stripe. Unfortunately the microwave signal suppression from terminal~1 to terminal~3 is only about 10~dB (as shown in supplementary figure S1) leading to a cross talk ac signal amplitude on terminal~3 of about 15~dBm. This signal is 2 orders of magnitude larger than the  expected ac-ISHE signal. For this reason it is  compensated by a reference signal using a power combiner where amplitude and phase of the reference signal can be adjusted to almost fully compensate the direct cross talk between terminals~1 and 3. The resulting signal has a magnitude of only a few mV allowing the detection of the ISHE by a power meter (detection scheme 1) or by a rectifying diode and a lock-in amplifier (detection scheme 2). For lock-in detection the static magnetic field is modulated with an amplitude of 0.2~mT. The lock-in signal is converted into the ac voltage amplitude at terminal 3 using the  power to voltage conversion characteristics  of the Schottky detector diode.


First the dynamic properties of the bilayer devices are studied by frequency dependent FMR measurements. For these measurements in-plane excitation is used and the magnetic field is applied along the $x$-axis ($\phi_{\rm{H}}=90^{\circ}$). The results are summarized in Fig.~1(c) where a typical FMR spectrum obtained  at a microwave frequency of 8~GHz is shown as the upper left inset. The FMR field $\rm{H_{r}}$ and line width $\Delta\rm{H}$ are extracted from the spectra as a function of frequency.  The frequency dependence of $\rm{H_{r}}$ can be well reproduced by a Kittel fit with effective magnetization $\mu_0M_{\rm{s}}=0.9~\rm{T}$. $\Delta\rm{H}$ is strictly proportional to the microwave frequency, and the Gilbert damping constant determined from the slope of $\Delta\rm{H}(f)$ is $\alpha=0.012$, which is enhanced compared to $\alpha=0.008$ for a reference Ni$_{80}$Fe$_{20}$ layer, due to spin pumping \cite{Tserkovnyak-PRL02, Mizukami-JMMM01}.


\begin{figure}[!ht]
\center \includegraphics[width=0.5\textwidth]{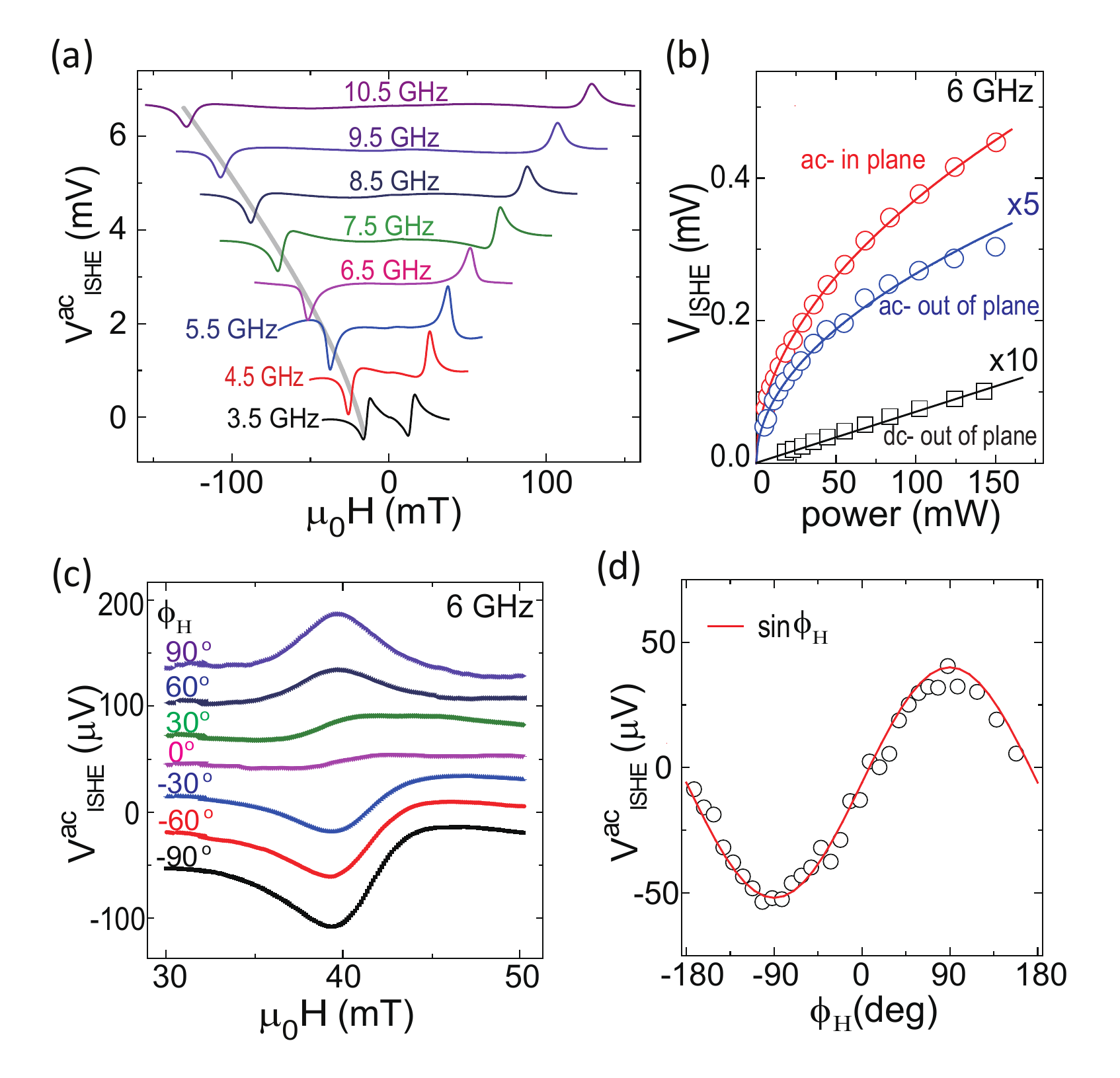} \caption
{\label{fig2}\textbf{The frequency, power and angular dependence of the ac-ISHE signals.} (a) The ac-ISHE voltages measured by the lock-in amplifier with microwave frequency from 3.5 to 10.5 GHz and in-plane excitation. (b) The microwave power ($P$) dependence of $\rm{V^{ac}_{ISHE}}$ (in-plane and out-of-plane excitation) and $\rm{V^{dc}_{ISHE}}$ (out-of-plane excitation) at 6~GHz, the ac- and dc- signals measured with out-of-plane excitation are multiplied by 5 and 10 for comparison. The solid lines are fits to $\sqrt{P}$ and $P$ for ac- and dc-ISHE, respectively. (c) ac-ISHE measured at 6 GHz out-of-plane excitation with different field angles $\phi_H$ from $-90^{\circ}$ to $90^{\circ}$. (d) Angular dependence of the intensity of $\rm{V^{ac}_{ISHE}}$. } \label{fig2}
\end{figure}

Typical signals of the ac-ISHE ($\rm{V^{ac}_{ISHE}}$) measured on a Ni$_{80}$Fe$_{20}$/Pt stripe at 8~GHz using in-plane excitation are shown in  Fig.~1(d). The top spectrum spectrum (red line) is the amplitude of the ac voltage along the Ni$_{80}$Fe$_{20}$/Pt stripe measured directly with a microwave power meter (detection scheme 1); as outlined in Fig.~1(b). At the resonance field, a step like feature  with an amplitude of 450 $\mu$V is observed. This signal is attributed to the ac-ISHE. The bottom  spectrum (blue line) is the ac-ISHE signal measured by field modulation and lock-in amplification (detection scheme 2). This spectrum was converted into the voltage $\rm{V^{ac}_{ISHE}}$ by numerical integration. Line shape and amplitude are in agreement with the spectrum observed by the power meter however with significantly improved signal to noise ratio.

In the following the line shape, frequency, angular and power dependence of the observed ac-ISHE signal will be examined in detail.

First we would like to address the shape of the ac-ISHE signals. The signals we measure are a superposition of a field independent microwave electric field (cross talk between terminals 1 and 3) and the actual ac-ISHE signal. The antisymmetric line shape observed in Fig.~1(d) is a consequence of this superposition. Since the relative phase shift $\Phi_{0}$ between the electric cross talk and the ac-ISHE signal is frequency and sample dependent any line shape (symmetric to antisymmetric) can result. This is demonstrated by recording  $\rm{V^{ac}_{ISHE}}$ spectra at frequencies between 3.5 and 10.5 GHz (from bottom to top) shown in Fig.~2(a). As a function of microwave frequency the ac-ISHE signals are observed at the negative and positive resonance fields of FMR, indicated by the dashed gray line. The shapes of the resonance in $\rm{V^{ac}_{ISHE}}$ can be peaks, dips, or fully anti-symmetric signals depending on the microwave frequency and device. The line shape of these spectra can be well explained by the superposition  of two ac signals.  A numerical simulation of the sum of  $\rm{V^{ac}_{ISHE}}$  and $\rm{V^{bac}}$ for different phase shifts $\Phi_{0}$ between these two signals \cite{wirthmann2010direct} is shown in the Supplementary figure S2.

Fig.~2(b) shows the rf-power dependence of $\rm V_{ISHE}$ at 6~GHz. The red dots and blue squares are for the ac- and dc-ISHE amplitudes respectively.  $\rm{V^{dc}_{ISHE}}$ and $\rm{V^{dc}_{ISHE}}$ are measured on different devices with in- and out-of-plane excitation field, respectively.  $\rm{V^{dc}_{ISHE}}$ is proportional to the rf-power $P$ \cite{azevedo2011spin,costache-PRL2006}, $\rm{V^{ac}_{ISHE}}$ on the other hand scales with $\sqrt{P}$. This power dependence will be discussed below.

The angular dependence of $\rm{V^{ac}_{ISHE}}$ measured at 6~GHz is shown in  Fig.~2(c) and (d). For this experiment out-of-plane excitation is used and a rotatable magnetic field $H$ is applied in the $x-y$ plane, thus the magnetic excitation and in this way the spin pumping process do not depend on the in-plane field angle $\phi_{\rm{H}}$.
The spectra for $\phi_{\rm{H}}$ between 90 to -90$^{\circ}$ (from top to bottom) are shown in Fig.~2(c). The spectrum at $\phi_{\rm{H}}=90^{\circ}$ ($H$ applied along the stripe) shows a symmetric line shape, and its intensity decreases monotonically to zero when $\phi_{\rm{H}}$ is 0$^{\circ}$ ($H$ perpendicular to the stripe); for even smaller angles the signal reverses. The amplitude of $\rm{V^{ac}_{ISHE}}$ as a function of $\phi_{\rm{H}}$ is shown in Fig.~2(d) and can be well fitted to a sine function, as expected from Eq.~\ref{she}. At $\phi_{\rm{H}}=0^{\circ}$, since the ac spin current spin direction ${\bm \sigma}$ is rotating in the $x-z$ plane, the ac-ISHE voltage is generated along the $\emph{y}$ direction, leading to a vanishing voltage along the $x$ direction (along the stripe).


In the following we compare the amplitudes of the dc- and ac-ISHE signals measured in the same device. For  the dc-ISHE measurements the voltage is measured by connecting a nanovoltmeter to terminal~3 of the sample. In Fig.~3(a), the top (red) and bottom (black) spectra are the ac- and dc-ISHE voltages measured at 6~GHz with out-of-plane excitation. One can clearly see that the ac-ISHE signal is much larger than the dc-ISHE signal. For the measurement of the ac-ISHE the applied field is oriented at $\phi_{\rm{H}}$ of 90$^{\circ}$ while for the dc-ISHE  $\phi_{\rm{H}}=0^{\circ}$ is used, cf.~Eq.~\ref{she} and Fig.~1(a).  For the measurements in Fig.~3(a) we obtain a magnitude  of $\rm{V^{ac}_{ISHE}}$ and $\rm{V^{dc}_{ISHE}}$ of 60 and 10~$\mu$V, respectively.

Theoretically one can derive the following expressions for the peak amplitudes \cite{Martin2013SHE} (and supplementary discussion part 1):
\begin{eqnarray}  \label{ac-ishe-eq}
\rm{V^{ac}_{ISHE}} = \it \alpha_{\rm{SH}}\frac{e}{\sigma_{\rm{NM}}}\frac{\rm{1}}{\rm{2\pi}M_{\rm{S}}}\frac{\lambda_{\rm{sd}}}{t_{\rm{NM}}}\it{l}
\tanh(\frac{t_{NM}}{\rm 2\lambda_{sd}})\notag\\\times\gSpinMix\omega{}h_{z}\Im(\chi^{res}_{zz})\cos{(\omega{}t)}
\end{eqnarray}
\begin{eqnarray}  \label{dc-ishe-eq}
\rm{V^{dc}_{ISHE}} = \it
\alpha_{\rm{SH}}\frac{e}{\sigma_{\rm{NM}}}\frac{\rm{1}}{\rm{2\pi}M^2_{\rm{S}}}\frac{\lambda_{\rm{sd}}}{t_{\rm{NM}}}\it{l} \tanh(\frac{t_{NM}}{\rm 2\lambda_{sd}})\notag\\\times\gSpinMix\omega{}h^2_{z}\Im(\chi^{res}_{zz})\chi^{res}_{yz}
\end{eqnarray}
here $\alpha_{\rm{SH}}$ and $\lambda_{\rm{sd}}$ are the spin Hall angle and spin diffusion length of NM, $l$ is the length of the stripe, and $\chi^{res}_{yz}$ and $\chi^{res}_{zz}$ are the in and out-of-plane susceptibilities at FMR. $\gSpinMix$ is the spin mixing conductance, $\sigma_{\rm{NM}}$ is the conductivity of the bilayer and $t_{\rm{NM}}$ is the thicknesses of the normal metal (Pt) layer.  Since the dc-ISHE scales with $h_z^2$ one expects $\rm{V^{dc}_{ISHE}}$ to scale linear with the microwave power $P$ \cite{azevedo2011spin,costache-PRL2006}, while for ac-ISHE a scaling behavior with $P^{1/2}$ is expected. This behavior is perfectly reproduced in our experiment as shown in Fig.~2(b). Furthermore, the expected ratio of the amplitudes of the ac- and dc-ISHE voltages is given by
\begin{eqnarray}
\frac{\rm{V^{ac}_{ISHE}}}{\rm{V^{dc}_{ISHE}}} = \frac{M_{\rm{S}}}{\chi^{res}_{yz}h_{z}}
\end{eqnarray}
 This ratio can be easily calculated for parameters that apply to Fig.~3(a): $f=6$~GHz ($\leadsto H_ {\rm FMR}=45$~mT), using $\mu_0 M_{\rm{S}}= 0.9$~T, $\chi^{res}_{yz}=61.5$ and $\mu_0 h_{z}=0.4$~mT \cite{mosendz2010quantifying} results in $\rm V^{ac}_{ISHE}/V^{dc}_{ISHE}=40$. Experimentally we only observe  $\rm V^{ac}_{ISHE}/V^{dc}_{ISHE}=6$, however one needs to consider the poor transmission of the rf-signal into the 50~$\Omega$ terminal. In the case of out-of-plane excitation the bilayer stripe in the gap of the CPW can be considered as a waveguide with a characteristic impedance of 250~$\Omega$. The resistance mismatch between this waveguide and the 50~$\Omega$ terminal leads to a transmission of only 33\% of the signal as can be calculated from the voltage standing wave ratio $T=1-\frac{Z_0-Z_1}{Z_0+Z_1}$. This implies that the $\rm V^{ac}_{ISHE}/V^{dc}_{ISHE}=6$ is in fact 3 time larger on the sample. Even in this case the observed $\rm V^{ac}_{ISHE}$ is still half of expected. Considering the poor microwave properties of the  bonding wires that are used to connect the magnetic bilayer (ac signal source) and terminal 3 and the Ohmic losses in the bilayer stripe, a loss of 1/2 of the signal at 6~GHz is reasonable. This means that also the absolute amplitude we observe is roughly in agreement with theory \cite{Jiao-PRL13}. Using published values for the spin Hall angle $\alpha_{\rm{SH}}=0.08$ \cite{liu2011spin} and $\lambda_{\rm{sd}}=1.4$~nm \cite{liu2011spin,liu2011review}, the ISHE voltages at resonance (6~GHz) can be calculated as $\rm V^{ac}_{ISHE}=0.4$~mV and $\rm V^{dc}_{ISHE}=10$~$\mu$V, respectively.

\begin{figure}[!ht]
\center \includegraphics[width=0.4\textwidth]{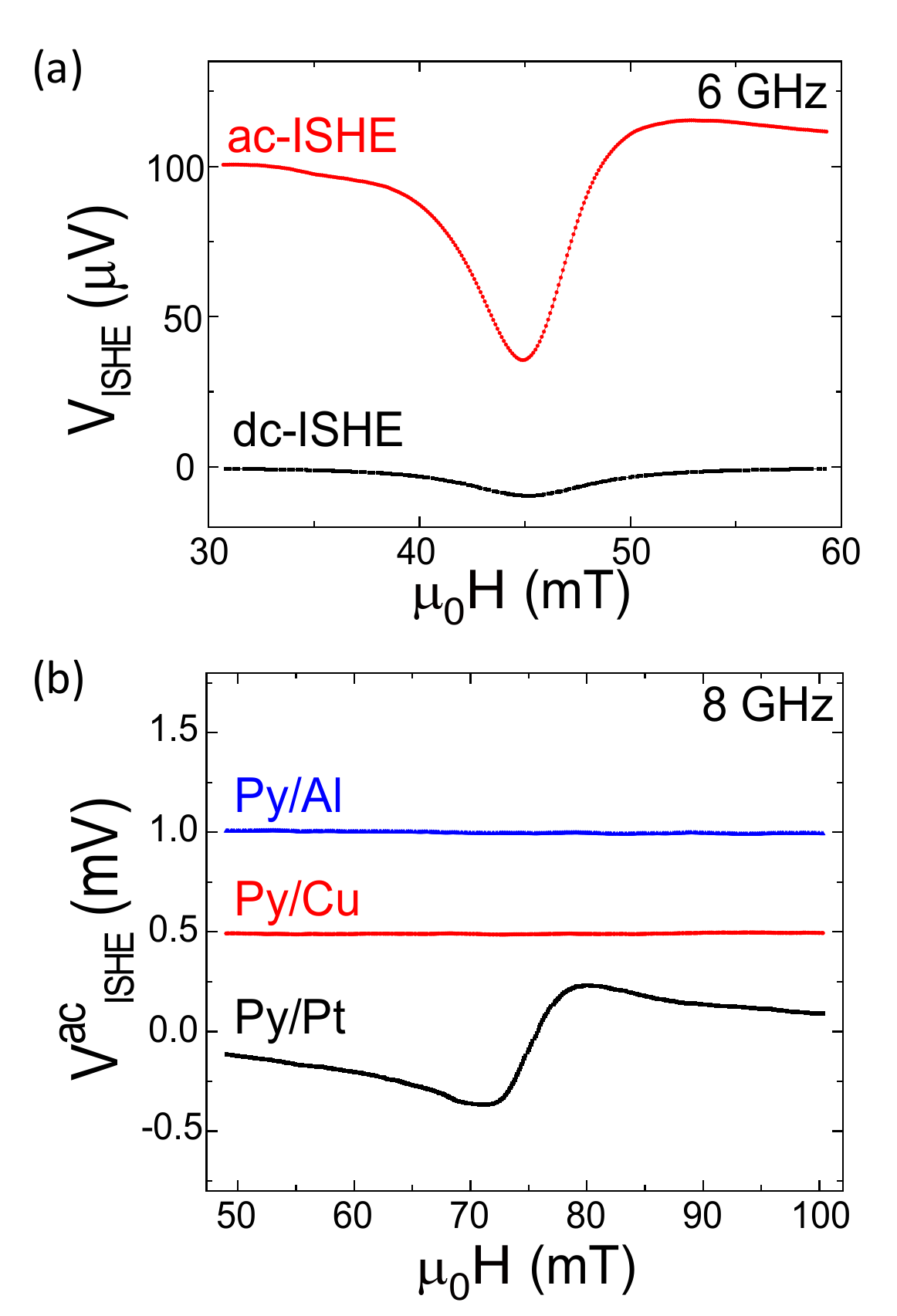} \caption
{\label{fig3}\textbf{Comparison of the ac and dc-ISHE amplitude and material dependence.} (a) Comparison of the ac- and dc-ISHE voltages for the same device measured at 6~GHz in the out-of-plane excitation configuration. The ac-ISHE voltage is about 30 times larger than the dc one. (b) Comparison of the ac-ISHE signals for Ni$_{80}$Fe$_{20}$/Al, Ni$_{80}$Fe$_{20}$/Cu and Ni$_{80}$Fe$_{20}$/Pt measured at 8 GHz. Only for Ni$_{80}$Fe$_{20}$/Pt the ac-ISHE signal observed at FMR. } \label{fig3}
\end{figure}
A similar analysis can be performed with the signal amplitudes shown in Fig.~2(a). E.g.~for FMR at 10~GHz one obtains the following parameters: $\chi^{res}_{yz}=56.5$ (due to the in-plane excitation $\chi^{res}_{yz}$ has to be used instead of $\chi^{res}_{zz}$) and $\mu_0 h_{y}=0.27$~mT. From this and Eq.~\ref{ac-ishe-eq} an amplitude of 3.1~mV is expected. Again the waveguide properties of the Ni$_{80}$Fe$_{20}$/Pt stack on the gold waveguide need to be considered. As discussed in the supplementary discussion part 2, this configuration is equivalent to a microstrip waveguide with a characteristic impedance $Z_0=480~\Omega$. One expects a transmission of only 18\% into $Z_1=50$~$\Omega$ using the voltage standing wave ratio (supplementary discussion part 2). Again due to ohmic losses in the bilayer and poor microwave transmission of the bonding wires one finds approximately 1/2 of the expected amplitude (0.6~mV), see Fig.~2(a).

Additional parasitic signals may in principle be generated by the anisotropic magneto-resistance (AMR) in the Ni$_{80}$Fe$_{20}$ layer and also contribute to ac- and/or dc- voltages along the Ni$_{80}$Fe$_{20}$/Pt stripe. Via the non-linear rectification effect, a combination of the inductively coupled rf current in the stripe and a time dependent modulation of the resistance due to the AMR effect, an ac-voltage with twice the frequency and a dc-voltage may also be generated by mixing. In dc-ISHE experiments it is difficult to distinguish the ISHE-signal from the AMR voltage \cite{bai2013distinguishing,Martin2013SHE}. In contrast to this for the ac-ISHE measurements, AMR signals can only occur  at $2\omega$ and are easily suppressed using band pass filter (cf.~Fig.~1(b)). In our experiments we find no evidence of an AMR contribution to the signal, even when the band pass filter is omitted.

In order to further confirm that the observed signal is indeed a consequence of the inverse spin Hall effect in Pt, we compare samples with different capping layer materials. Devices with the following layer stacks were prepared: Ni$_{80}$Fe$_{20}$/Pt, Ni$_{80}$Fe$_{20}$/Al, and Ni$_{80}$Fe$_{20}$/Cu. The corresponding measurements performed at 8~GHz are shown in Fig.~3(b). For Al and Cu, it is well accepted that the spin Hall effect is very small due to the weak spin-orbit interaction \cite{Valenzuela-Nature06,Niimi-PRL11}. For otherwise identical experimental conditions the spectra are shown in Fig.~3(b). Only for the Ni$_{80}$Fe$_{20}$/Pt device the voltage signals are observed at the ferromagnetic resonance field. This material dependence and the fact that the angular dependence, line shape, and magnitude are in line with the theory of the ac-ISHE indicate that our signals are indeed a consequence of the spin currents generated by spin pumping and detected by the inverse spin Hall effect.


In summary, we present the first experimental demonstration of the ac-ISHE due to spin pumping at FMR. We demonstrate $\rm{V^{ac}_{ISHE}}$ signals with amplitudes of up to a few mV. The direct comparison of the ac- and dc- ISHE voltage on the same device for out-of-plane excitation shows that $\rm{V^{ac}_{ISHE}}$ is approximately 6 times larger than $\rm{V^{dc}_{ISHE}}$ despite the fact that our experiment can only detect 20\% of ac-ISHE signal. The large ac-ISHE voltages indicate the presence of large rf spin currents in agreement with the theory of spin pumping. Such spin currents and their detection via ISHE may be very useful for the development of ac spintronic devices.

\section{Methods}
The bilayer stripes are prepared by optical lithography, magnetron sputter deposition, and lift-off techniques on  semi-insulating GaAs substrates. Subsequently the CPW and the electrical contacts are fabricated by optical lithography using gold metallization. A thick alumina layer (80~nm $\rm{Al_{2}O_{3}}$) deposited by atomic layer deposition is used to insulate the Ni$_{80}$Fe$_{20}$/Pt stripe and its contact electrodes from the CPW.  The typical resistance of the Ni$_{80}$Fe$_{20}$/Pt stripes is around 1400~$\Omega$. All measurements are performed at room temperature with a nominally constant input microwave power of 320~mW (25~dBm).

The authors would like to acknowledge financial support from the German research foundation (DFG) through programs SFB 689 and SPP 1538 and from the European Research Council (ERC) through starting grant ECOMAGICS. D.W. would like to acknowledge a stipend from the Alexander von Humboldt foundation (AvH).

\bibliography{references}

\end{document}